\documentclass[12pt,psfig]{article}
\usepackage{amssymb}
\usepackage{amsmath}
\usepackage{epsfig}
\makeatletter
\def\@cite#1#2{{[{#1}]\if@tempswa\typeout
{IJCGA warning: optional citation argument
ignored: `#2'} \fi}}
\newcount\@tempcntc
\def\@citex[#1]#2{\if@filesw\immediate\write\@auxout{\string\citation{#2}}\fi
  \@tempcnta\z@\@tempcntb\m@ne\def\@citea{}\@cite{\@for\@citeb:=#2\do
            {\@ifundefined
       {b@\@citeb}{\@citeo\@tempcntb\m@ne\@citea\def\@citea{,}{\bf
?}\@warning
       {Citation `\@citeb' on page \thepage \space undefined}}%
    {\setbox\z@\hbox{\global\@tempcntc0\csname b@\@citeb\endcsname\relax}%
     \ifnum\@tempcntc=\z@ \@citeo\@tempcntb\m@ne
       \@citea\def\@citea{,}\hbox{\csname b@\@citeb\endcsname}%
     \else
      \advance\@tempcntb\@ne
      \ifnum\@tempcntb=\@tempcntc
      \else\advance\@tempcntb\m@ne\@citeo
      \@tempcnta\@tempcntc\@tempcntb\@tempcntc\fi\fi}}\@citeo}{#1}}
\def\@citeo{\ifnum\@tempcnta>\@tempcntb\else\@citea\def\@citea{,}%
  \ifnum\@tempcnta=\@tempcntb\the\@tempcnta\else
   {\advance\@tempcnta\@ne\ifnum\@tempcnta=\@tempcntb \else
\def\@citea{--}\fi
    \advance\@tempcnta\m@ne\the\@tempcnta\@citea\the\@tempcntb}\fi\fi}
\makeatother

\newcommand{\gsim}{\lower.7ex\hbox{$\;\stackrel{\textstyle>}{\sim}\;$}}
\newcommand{\lsim}{\lower.7ex\hbox{$\;\stackrel{\textstyle<}{\sim}\;$}}
\newcommand{\be}{\begin{equation}}
\newcommand{\ee}{\end{equation}}
\newcommand{\bea}{\begin{eqnarray}}
\newcommand{\eea}{\end{eqnarray}}

\usepackage{amssymb}

\def\baselinestretch{1}
\begin{document}
%%%%%%%%%%%%%%%%%%%%%%%%%%% subequations.sty %%%%%%%%%%%%%%%%%%%%%%%%
\catcode`@=11
\newtoks\@stequation
\def\subequations{\refstepcounter{equation}%
\edef\@savedequation{\the\c@equation}%
  \@stequation=\expandafter{\theequation}%   %only want \theequation
  \edef\@savedtheequation{\the\@stequation}% % expanded once
  \edef\oldtheequation{\theequation}%
  \setcounter{equation}{0}%
  \def\theequation{\oldtheequation\alph{equation}}}
\def\endsubequations{\setcounter{equation}{\@savedequation}%
  \@stequation=\expandafter{\@savedtheequation}%
  \edef\theequation{\the\@stequation}\global\@ignoretrue

\noindent}
\catcode`@=12
%%%%%%%%%%%%%%%%%%%%%%%%%%%%%%%%%%%%%%%%%%%%%%%%%%%%%%%%%%%%%%%%%%%%%
\begin{titlepage}
\title{{\bf Dark Energy, scalar-curvature couplings and a critical acceleration scale
}}
\vskip2in

\author{
{\bf Ignacio Navarro\footnote{\baselineskip=16pt E-mail: {\tt
ignacio.navarro@cern.ch}}}\\
%\hspace{3cm}
{\small CERN, Theory Division, CH-1211 Geneva 23, Switzerland.}\\
}

\date{}
\maketitle
\def\baselinestretch{1.15}

\begin{abstract}
\noindent We study the effects of coupling a cosmologically
rolling scalar field to higher order curvature terms. We show that
when the strong coupling scale of the theory is on the
$10^{-3}-10^{-1}eV$ range, the model passes all experimental
bounds on the existence of fifth forces even if the field has a
mass of the order of the Hubble scale in vacuum and non-suppressed
couplings to SM fields. The reason is that the coupling to certain
curvature invariant acts as an effective mass that grows in
regions of large curvature. This prevents the field from rolling
down its potential near sources and makes its effects on
fifth-force search experiments performed in the laboratory to be
observable only at the sub-mm scale. We obtain the static
spherically symmetric solutions of the theory and show that a
long-range force appears but it is turned on only below a fixed
Newtonian acceleration scale of the order of the Hubble constant.
We comment on the possibility of using this feature of the model
to alleviate the CDM small scale crisis and on its possible
relation to MOND.

\end{abstract}

\vspace{3cm}

\vskip-21.5cm \rightline{} \rightline{CERN-PH-TH/2007-155}
\end{titlepage}
%%%%%%%%%%%%%%%%%%%%%%%%%%%%%%%%%%%%%%%%%%%%%%%%%%%%%%%%%%%%%%%%%%%
\setcounter{footnote}{0} \setcounter{page}{1}

\baselineskip=20pt

If we want to obtain a scalar field model of dynamical Dark Energy
(DE) its mass has to be of the order of the Hubble scale\footnote{We will use natural units throughout the paper so that $c$ and $\hbar$ are set to one.} ($H_0
\sim 10^{-33}eV$) so that it evolves in the appropriate
time-scale. This, of course, raises the question of how to avoid
conflicts with the stringent limits on the (non) existence of
extra long range forces coming from laboratory experiments and
Solar System observations. Even if the scalar field couples to
ordinary matter only with non-renormalizable interactions
suppressed by the Planck scale, laboratory limits imply that its
mass should be larger than $10^{-3}eV$ \cite{Adelberger:2006dh}.
Several mechanisms such as approximate global symmetries to
suppress its couplings \cite{Carroll:1998zi} or an
environment-dependent effective potential and mass (the so-called
chameleon mechanism) \cite{Khoury:2003aq,Kaloper:2007gq} have been
proposed to reconcile a cosmologically rolling scalar field with
the negative results of fifth-force search experiments. In this
letter we consider the effect of coupling the quintessence field
to higher order curvature invariants. We will see that such
couplings provide a new way to suppress the effects of the force
mediated by the scalar field at the Solar System or laboratory
level and, at the same time, offer a range of unique observational
signatures detectable in laboratory experiments or astrophysical
observations. Specifically, the coupling we will be interested in
is \be \Delta {\cal L}=
\frac{\phi^2}{2\Lambda^2}R_{\mu\nu\lambda\sigma}R^{\mu\nu\lambda\sigma},\label{C1}
\ee where $\phi$ is the ultra-light field,
$R^{\mu}_{\nu\lambda\sigma}$ the Riemann curvature tensor and
$\Lambda$ some mass scale. Its origin could be in a coupling of
the scalar field with the Gauss-Bonnet combination, like in the DE
models of \cite{Nojiri:2005vv}, but since the scalar curvature or
Ricci tensor part of such a coupling would not play any role in
what follows we omit them for simplicity. And differently from the
models in \cite{Nojiri:2005vv} we will not use this kind of
coupling to drive the acceleration. We will assume that the field
$\phi$ has an appropriate quintessence-like potential to drive the
acceleration and this coupling can be simply interpreted as a
curvature-dependent mass for the scalar field. Its effect will be
to keep the field firmly locked at $\phi=0$ in regions of large
curvature, $e.g.$ near sources. However, on a cosmological scale
the field will have a small mass at late times and will be able to
roll down its potential. But for this to happen we must require
that the field mass obtained from this operator is small enough
today ($\Delta m_s^2 \lesssim H_0^2$). On a cosmological scale the
operator above will translate into a mass of order \be \Delta
m_s^2 \sim \frac{H_0^4}{\Lambda^2} \ee so the scale $\Lambda$ has
to satisfy $\Lambda \gtrsim H_0\sim 10^{-33}eV$ for the field to
be able to roll cosmologically. On the other hand if we want the
theory to pass the laboratory tests even if the field has non-suppressed couplings to matter we need its mass to be $m_s \gtrsim
10^{-3}eV$ locally. In the neighborhood of a massive object like
the Earth the Kretchmann scalar reads
$R_{\mu\nu\lambda\sigma}R^{\mu\nu\lambda\sigma} = 48\frac{(G M)^2
}{r^6}$, where $G$ is Newton's constant and $M$ is the mass of the
object, so that near sources the field gets a mass \be m_s^2
\simeq 48 \left(\frac{G M}{r^3\Lambda}\right)^2. \ee Plugging in
this expression the Earth's mass and radius and imposing that this
mass is larger than $10^{-3}eV$ we get the bound\footnote{Over the
distance scales associated with laboratory experiments we can take
the field mass as constant. And as long as the density of our
probe is less than the Earth's, the Earth's gravitational field
will give the dominant contribution to the scalar mass.} $\Lambda
\lesssim 10^{-31}eV$. So we see that these constraints leave open
just a small window for the scale $\Lambda$ of two orders of
magnitude. And notice that this window is in an extremely small
mass range of the order of the Hubble scale, $10^{-33}-10^{-31}
eV$, so one might wonder whether this theory could make sense as
an effective quantum field theory at all. But to obtain the
correct strong coupling scale of the theory we have to take into
account the proper (canonical) normalization of the fluctuations
of the metric. It is easy to see that the coupling (\ref{C1}), in
a weak field expansion
  in flat space in powers of the fluctuations of the metric
will give rise to couplings like \be
\frac{\phi^2}{\Lambda^2}R_{\mu\nu\lambda\sigma}R^{\mu\nu\lambda\sigma}
= \frac{\phi^2 \partial^2 h^{(c)} \partial^2 h^{(c)}} {\Lambda^2
M_p^2} + {\cal
O}\left(\frac{h^{(c)}(\partial^2h^{(c)})^2}{\Lambda^2
M_p^3}\right) \ee where we have suppressed the indices, and
$h^{(c)}$ represents the canonically normalized metric
fluctuations: $g_{\mu \nu} = \eta_{\mu\nu} +
h^{(c)}_{\mu\nu}/M_p$. We see then that the strong coupling scale
in this theory is $\Lambda_s \simeq \left(\Lambda
M_p\right)^{1/2}$ that will be in the range $10^{-3}-10^{-1} eV$.
So our scalar field will be strongly coupled to gravity, with a
strong coupling scale close to the vacuum energy scale\footnote{In
fact it has been argued in \cite{Beane:1997it} that taking a
cut-off in our effective gravitational theory at the vacuum energy
scale might be a way to understand the smallness of the
cosmological constant (see however \cite{Polchinski:2006gy}).}.
And notice also that the mass of the field locally in the Earth is
pushed close to, or might even get above, the cut-off scale of the
theory. So, depending on the chosen parameters, we might get out
of the range of validity of the theory when computing its
implications for laboratory searches of fifth forces. But in any
case we can expect that, like in the chameleon models of
\cite{Khoury:2003aq}, the mass of the field will decrease and the
range of the force mediated by this field will increase if an
experiment is performed in space offering a very characteristic
experimental signature.

But even if the field is massive close to sources it is nearly
massless in regions of small curvature. It is therefore natural to
ask what is the effect of the coupling (\ref{C1}) for long range
forces. In order to gain some intuition about the
  expected behavior of the spherically symmetric solutions we can present the following argument. We expect that generically a massive boson will mediate a
  fifth force appreciable when the distance to the source is $smaller$
  than its
  inverse mass. In these situations the effect of the mass term in the equation of motion is
  negligible as compared with the terms coming from the Laplacian ($i.e.$ momentum
  scale). We can approximate then the equation to solve by $\nabla^2 \phi=\phi^{\prime \prime}+2\phi^{\prime}/r =0$, yielding the usual $\phi \propto 1/r$ profile characteristic
  massless fields. But for longer distances the mass term is not negligible anymore with respect to the
  derivative terms in the equations and this
  suppresses exponentially the solution. This well-known Yukawa exponential suppression ($\phi \propto e^{-m_sr}/r$) is the reason why massive particles mediate only
  short range forces, significant only when $r<m_s^{-1}$. But now our effective mass term reads $\Delta m_s^2 \sim 48\left(\frac{GM}{r^3\Lambda}\right)^2$ in
  the Schwarzschild geometry, so it
depends on the distance to the source. And according to the
previous argument, if
  the coupling above gives the dominant contribution to the mass we
  can expect an appreciable effect only whenever
\be r<m_s(r)^{-1}\sim \frac{\Lambda r^3}{GM}. \ee Since now the
mass grows at short distance this condition means that the force
would only appear at $long$ distances, $r>r_c$, where $r_c$,
defined as \be \frac{GM}{r_c^2} \equiv \Lambda, \label{A1} \ee
corresponds to a fixed gravitational acceleration\footnote{This
dependence of the mass with the distance yielding a fixed
Newtonian acceleration as the threshold for the appearance of new
effects is the same one finds for the scalar mode of the metric in
the purely gravitational model of \cite{Navarro:2005ux}.} $a_c=
  \Lambda$. We will confirm this result by explicitly obtaining
  the static spherically symmetric solutions to the
equation of motion for the field $\phi$ with this $r$-dependent mass
term. The fact that the force mediated by this field appears only
below a fixed Newtonian acceleration scale makes this model
interesting from a phenomenological point of view, since it can
produce the emergence of a critical
  acceleration in dark matter (DM) halos helping with the problems of conventional cold dark matter
  (CDM) models at small scales.

Let us present here a brief review of these problems, that will
serve as an extra motivation for the kind of phenomenology that
the present model can offer. The reader uninterested in this
discussion can skip directly to the next paragraph. It can be said
that the comparison of the structure of galaxies and DM halos
formed in numerical simulations (within the $\Lambda$CDM paradigm)
with their real observational counterparts is far from
satisfactory. The main discrepancies come from the smallest
scales: the simulations yield too much DM density in the
central part of collapsed structures and too many
sub-structures within DM halos (see $e.g.$
\cite{Ostriker:2003qj,Spergel:1999mh} and references therein).
This is the main cause of what is sometimes called
the CDM ``small scale crisis'', and it has prompted the
consideration of other flavors of DM, such as self-interacting \cite{Spergel:1999mh}, warm \cite{Bode:2000gq} or decaying DM \cite{Cembranos:2005us}. However it
is unclear if a DM self-interaction is compatible with the
structure at other scales like cluster halos
\cite{Yoshida:2000bx}. Also, the possibility of having warm DM is
severely constrained by observations of Ly-$\alpha$ forest
\cite{Seljak:2006qw} and it does not seem compatible with the evolved structure seen at high redshifts,
since it acts delaying structure-formation \cite{Haiman:2001dg}
(and the same constraints limit decaying DM
scenarios \cite{Jedamzik:2005sx}). But on top of this, there are
certain systematic features of galactic dynamics that seem hard to
explain in any conventional DM scenario. In particular the onset
of the mass discrepancy or need for DM seems to be tightly
correlated with a critical acceleration scale\footnote{This empirical fact underlies the success of MOND, a phenomenological
modification of Newton's potential for small accelerations
that is able to predict with remarkable accuracy the
rotation curve of spiral galaxies from its visible matter content
\cite{Sanders:2002pf}.}, such that when the
Newtonian acceleration produced by any given source is above it,
there is no evidence for DM and when it is below we need to assume
the presence of unseen forms of matter to explain the dynamics
\cite{McGaugh:2004aw}. And this critical acceleration scale is very
small, curiously of the
order of the current Hubble constant. The above considerations, plus the cosmological constant problem, could be interpreted as hints of the existence of non-trivial
physics within the ``dark sector''\footnote{These motivations are of
  a theoretical nature for the
DE, but purely observational for DM.}. And the mechanism presented
here seems a good candidate to alter the physics of the infrared
domain of gravity in the right direction. It can explain the
appearance of a fixed acceleration scale in DM halos and has the
potential to save some of the contenders of CDM like warm or
decaying DM by modifying the physics of structure
formation because it can be seen as an enhancement of gravity in
low density regions.

We shall continue now exploring the implications of this kind of
interaction for long range forces. As usual we will assume a
coupling of this
  field to ordinary matter such that in the non-relativistic limit it
  can be parametrized as
\be
\Delta S_{int} =  \int d^4x \sqrt{g}\; \beta \frac{\phi}{M_p}\rho\label{coupling}
\ee
where $\rho$ is the matter energy density and $\beta$ a dimensionless
  parameter. The equation of motion for the scalar field reads then
\be \Box^2 \phi + \frac{Q}{\Lambda^2}\phi =
\beta\frac{\rho}{M_p}\label{E1} \ee where we have defined $Q$ as
the Kretchmann scalar $Q\equiv
R_{\mu\nu\lambda\sigma}R^{\mu\nu\lambda\sigma}$. We are neglecting
the potential for $\phi$ since we are assuming that
  we are in the regime where the coupling (\ref{C1}) gives the dominant
  contribution to the mass and the terms in this equation coming from
  the potential $V(\phi)$ are negligible.
To find a solution one has to specify $Q(x)$, $i.e$ the background
  geometry. In principle this should be done self-consistently, which
  means solving at the same time the coupled gravitational equations
  taking into account the back-reaction of the scalar field
  configuration in the background geometry. We will neglect now this backreaction and we will check later the conditions under which this is actually a good approximation. In this case the Kretchmann scalar outside a spherical mass distribution reads
  $Q=48\left(\frac{GM}{r^3}\right)^2$, so the equation that we have to solve in vacuum in order to find the static solutions is
\be
\frac{d^2 \phi}{dr^2} + \frac{2}{r}\frac{d\phi}{dr} -
  48\left(\frac{GM}{\Lambda r^3}\right)^2 \phi = 0.
\ee
The general solution to this equation takes the form
\be
\phi(r) = \frac{{\cal C}_1}{\sqrt{r}} I_{-1/4}(z) + \frac{{\cal C}_2}{\sqrt{r}} K_{1/4}(z)\label{S1}
\ee
where $I_{n}(z)$ and $K_{n}(z)$ are the modified Bessel functions of
  the first and second kind respectively. We have defined $z\equiv \frac{2\sqrt{3}GM}{\Lambda
  r^2}$ and the ${\cal C}_i$ are the integration constants. Notice
  that $z(r)$ is basically the ratio of the gravitational acceleration to
  the critical acceleration $a_c=\Lambda$. To
  determine the integration constants we have to satisfy the appropriate boundary conditions at infinity and in the source. As we said the field will go to zero as we approach the source\footnote{Actually, inside the source the equilibrium value for the field will be displaced from zero because the source acts as a tadpole through the coupling (\ref{coupling}), but the field will go to zero exponentially on a distance scale $\sim m_s^{-1}$ once we get out of the source.}. For large $z$ (or small $r$) we can approximate $I_{-1/4}(z)\propto
  e^{z}/z^{1/2}$ and $K_{1/4}(z)\propto
  e^{-z}/z^{1/2}$. So requiring that the solution is
  well behaved at small $r$ (remember that $z \propto 1/r^2$) we see that we are forced to take ${\cal C}_1=0$. Notice that these solutions are analogous to the familiar $\phi \propto e^{\pm m_sr}/r$ solutions for massive fields. Indeed, the heuristic argument presented before is backed by these exact solutions since $e^{-z}\sim e^{-m_s(r)r}$,
and $\phi \propto I_{-1/4}(z)/\sqrt{r}$ is the analogous to the
unphysical solution $\phi \propto e^{+m_sr}/r$ for massive fields.
But curiously the role of the boundary conditions at the source
and at infinity is reversed now with respect to the conventional
case of a field with constant mass. In the conventional case we
take the field to be asymptotically in its vacuum, say at
$\phi=0$, so the coefficient of the  $e^{+m_sr}/r$ solution is
taken to be zero. The coefficient of the $e^{-m_sr}/r$ term is
obtained by satisfying the boundary conditions at the source
giving the source mass as the constant of proportionality (the solution would be $\phi =
\beta \frac{M}{4\pi M_p}\frac{e^{-m_sr}}{r}$ for a coupling to matter like
the one in (\ref{coupling})). Now the field is locked at $\phi=0$
at short distances, forcing as we said ${\cal C}_1=0$, while we
allow it to be rolling at infinity, where it will have an
unspecified, time dependent v.e.v. that we will call
$\phi_\infty$. As $r\rightarrow \infty$ we get the limit \be
K_{1/4}(z)/\sqrt{r}\rightarrow 2\cdot 3^{-1/8}\Gamma
(5/4)\left(\frac{\Lambda}{GM}\right)^{1/4} \sim
1.58\left(\frac{\Lambda}{GM}\right)^{1/4},\ee where $\Gamma(x)$ is
the Euler gamma function. So the static\footnote{In reality the
solution is not static because we are considering time-dependent
asymptotic boundary conditions. But we are assuming here that the
relaxation time of the system is much smaller than the time-scale
associated with the change in boundary conditions ($\sim H_0^{-1}$), so this static
solution is a good approximation at any given time.} solution
taking into account this boundary condition at infinity will be
\be \phi(r) = \frac{\phi_\infty 3^{1/8}}{2\Gamma
(5/4)}\left(\frac{G
M}{\Lambda}\right)^{1/4}\frac{1}{\sqrt{r}}K_{1/4}(z) \simeq
\frac{\phi_\infty}{1.58}\left(\frac{G
M}{\Lambda}\right)^{1/4}\frac{1}{\sqrt{r}}K_{1/4}(z). \ee 
\begin{figure}[h]
  \begin{center}
    \epsfig{file=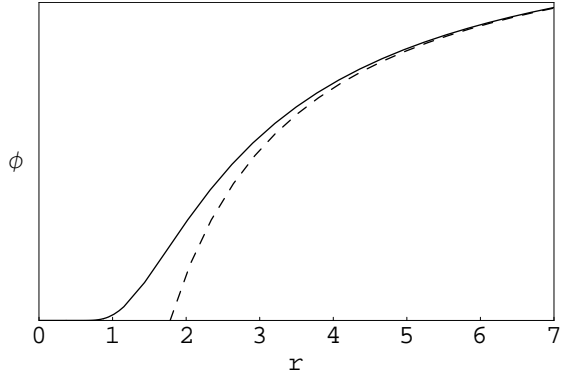,height=5cm}
\caption{The solution for the scalar field with $r_c=1$. The dashed line corresponds to the asymptotic $1/r$ dependence.}
\end{center}
\end{figure}
As we
said this solution is exponentially suppressed at small distances
(for $z>1$) but at large distances it asymptotes to $\phi_\infty$
with a $1/r$ profile as can be seen from the large distance
expansion 
\bea
\phi(r) & \simeq & \phi_\infty\left(1 + \frac{3^{1/4}\Gamma(-1/4)}{\Gamma(1/4)}\left(\frac{G M}{\Lambda}\right)^{1/2}\frac{1}{r} + {\cal O}\left(\frac{G^2 M^2}{\Lambda^2 r^4}\right)\right) \nonumber \\
& \simeq & \phi_\infty\left(1 -1.78 \left(\frac{G
M}{\Lambda}\right)^{1/2}\frac{1}{r} + {\cal O}\left(\frac{G^2
M^2}{\Lambda^2 r^4}\right)\right). \eea
 And notice that although
this solution depends on the source mass and asymptotic field value, it is independent of the
coupling of the field to matter, eq.(\ref{coupling}). We also see
how we recover the usual $1/r^2$ falloff in the force caused by
$\phi$ when $r\gg r_c$.
 In fig.1 we present a plot of this solution and in fig.2 a
  plot of its gradient, that would correspond to the force that a
  test particle would feel due to the profile of $\phi$ in this spherically symmetric
  solution. It is apparent how this force is switched on when $z\sim
  1$ (or $r \sim r_c$) and gets exponentially suppressed at smaller
  distances.

But we still have to check that the backreaction of the scalar field configuration on the background is actually negligible, so that taking $Q=48\left(\frac{GM}{r^3}\right)^2$ is a good approximation. Since the scalar field is exponentially suppressed for $r< r_c$, we
know that the backreaction is negligible in this regime and the
solution is valid. We will check now under which conditions neglecting
the backreaction for $r>r_c$ is justified.
This can be quantified by looking at the $\phi$-dependent terms in the
gravitational equations. The relevant terms in
the equations of motion for the metric read
\bea
M_p^2 \left(R_{\mu\nu} - \frac{1}{2}g_{\mu\nu}R\right)& = & T_{\mu\nu}^{(\phi)} + \frac{2}{\Lambda^2}
\nabla_\sigma
\nabla_\lambda \left(R_{(\mu \;\;\;\; \nu)}^{\;\;\; \sigma \lambda} \phi^2\right)\nonumber \\ &-& \frac{\phi^2}{\Lambda^2} R_{\sigma \lambda \gamma \mu}   R^{\sigma \lambda \gamma}_{\;\;\;\;\;\; \nu}
+ g_{\mu \nu}\frac{\phi^2}{4\Lambda^2} R_{\sigma \lambda \gamma \kappa}   R^{\sigma \lambda \gamma \kappa}
\eea
where $T_{\mu\nu}^{(\phi)}$ is the usual energy-momentum tensor for
$\phi$.
\begin{figure}[h]
  \begin{center}
    \epsfig{file=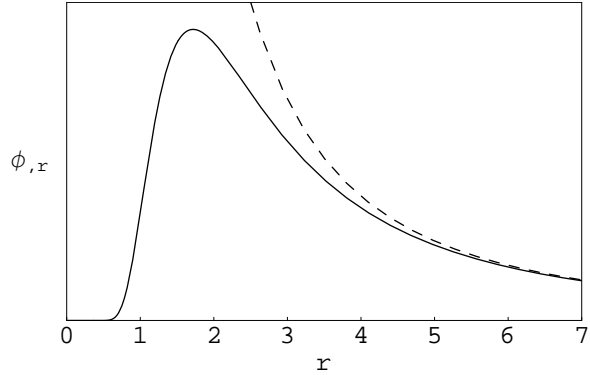,height=5cm}
\caption{The gradient of the scalar field ($i.e.$ the force) with $r_c=1$. The dashed line corresponds to the asymptotic $1/r^2$ dependence.}
\end{center}
\end{figure}
We can now interpret the right hand side of these equations as an
``effective energy-momentum density'', ${\cal
T}_{\mu\nu}^{(eff)}$, that will have an influence on the geometry.
So if the ``energy-momentum'' stored in this r.h.s. of the
Einstein equations is small compared to $M$ we can neglect its
influence on the background, while if it was bigger it can have a
significant impact on the long-distance geometry. We are
interested thus on the integration of this quantity as compared to
$M$ at any given radius. As mentioned before it is obvious that
for $r<r_c$ we can neglect ${\cal T}_{\mu\nu}^{(eff)}$  because
$\phi$ is exponentially suppressed. It is easy to check that
asymptotically in our solution, when $r\gg r_c$ and $\phi \sim
1/r$, the r.h.s. of the gravitational equations has the dependence
\be {\cal T}_{\mu\nu}^{(eff)} \sim  \frac{\phi_\infty^2  G^2
M^2}{\Lambda^3 r^7}. \ee In order to get an estimation of the
order of magnitude of the integration of ${\cal
T}_{\mu\nu}^{(eff)}$ we will approximate $\phi$ by this long
distance solution for $r>r_c$ and by zero for $r<r_c$. We get then
\be M^{(eff)} = \int\; d^3x {\cal T}_{00}^{(eff)} \sim \int
\frac{\phi_\infty^2 G^2 M^2}{\Lambda^3 r^7} r^2 dr \sim
\frac{\phi_\infty^2G^2M^2}{\Lambda^3 r_c^4} \sim
\frac{\phi_\infty^2}{\Lambda}. \ee Comparing this with $M$ we see
that in order to be able to neglect the backreaction we must have
$\phi_\infty \ll (\Lambda M)^{1/2}$.

It is also interesting to compare the usual Newtonian
gravitational force with the long distance force that will appear
when $r> r_c$ as a result of this profile for the scalar field.
Their ratio, for $r\gg r_c$, is \be \frac{\beta \phi(r)/M_p}{GM/r}
\sim \frac{\beta \phi_\infty}{(\Lambda M)^{1/2}}. \ee Notice that,
when $\beta$ is of order one, the condition that we need for the
backreaction of the scalar field on the geometry to be negligible is
precisely that this ratio is small. So if $\beta \sim 1$ and the
extra force due to the scalar becomes of the order of the
gravitational one we can not neglect the influence of the scalar
field configuration on the background geometry and our solution is
not valid\footnote{However, if $\beta \gg 1$ we can still have an
extra force stronger than gravity appearing at long distances
while the backreaction is small, so we can trust our solution.}.
And since $(\Lambda M)^{1/2} \ll M_p$ for any reasonable $M$, this
ratio will be very large unless $\phi_\infty \ll M_p$ as well, so
we can only use the solutions we have presented when the v.e.v. of
the field at infinity is small compared to the Planck mass. But
although the precise form of the solution might be different when
the backreaction is not negligible, we can still expect that the
solution will be suppressed for $r<r_c$ (since the backreaction is
negligible in this regime) and will recover a $\phi \propto 1/r$
profile at large distances, $r\gg r_c$ (since the whole mass term
for the scalar will be negligible in this regime and the field
will behave as effectively massless). Nevertheless, it would be
interesting to explore in more detail the behavior of the
solutions when the backreaction is not negligible. In this case
the solution for the metric will suffer an appreciable
modification (with respect to the Schwarzschild solution), and
this brings hopes that a model along these lines could be used to
build MOND-like modification of gravity that bypasses the need for
CDM. When the backreaction is negligible and the solution for the
metric is not significantly altered we can not hope to build a
realistic model that does not need DM because the effects of DM
are also felt in gravitational lensing, and this effect can not be
achieved without modifying the actual solution for the metric at
long distances (see $e.g.$ \cite{Bekenstein:2004ne}).

To conclude, we have explored the effects of coupling the
quintessence field to higher order curvature invariants and in
particular we have focused in a coupling that can be interpreted
as a curvature-dependent mass for the scalar field. On a
cosmological scale the field mass is small so it can roll down its
potential acting as dynamical Dark Energy. But as we approach
sources this mass grows, driving the field towards zero, and we
have found that if this coupling is strong enough it can shield
the Solar System from long range forces mediated by this scalar.
It is also worth mentioning that this effect also implies that no
time variation of fundamental constants associated with this
scalar field would be observed in local experiments\footnote{Or in
the early Universe, since the field mass would be much bigger than
the Hubble constant at early cosmological times and the field
would be settled in its high-curvature minimum at $\phi=0$.}.
However, in the places where the field is rolling, basically
inter-stellar or inter-galactic space, a time dependence of
fundamental constants could be observed. So this kind of model has
also the potential to reconcile the claims of the detection of a
time evolution of fundamental constants derived from quasar absorption spectra with conflicting bounds obtained locally, like those coming from the natural Oklo reactor (see \cite{Uzan:2002vq} for a
review). Furthermore, the model also offers the opportunity to be
tested in table-top laboratory experiments, and like the chameleon
models of \cite{Khoury:2003aq} boasts the characteristic signature of
mediating a force whose range would be larger on space, where the
local curvature is significantly lower. Needless to say, this
theory has many more implications for astrophysics because at long
distances an extra force will appear, becoming noticeable only
when the gravitational acceleration produced by sources fall below
a given scale. This feature makes the model interesting from a
phenomenological point of view, since it appears to have the right
properties to help with the problems of CDM at small scales. Interestingly, the strength of this force depends on the source's mass and asymptotic value of the field, but it is independent of the coupling of the field to matter.
However a more detailed study of the effects of this kind of
interaction in the formation and evolution of structures is beyond
the scope of the present paper.

\section*{Acknowledgments}

I would like to thank G. Efstathiou and K. Van Acoleyen for conversations.

\end{document}